\documentclass{epl}
\usepackage{amssymb,graphicx,psfrag,amsmath,cite}
\title{The angular dependent magnetoresistance in 
$\alpha$-(BEDT-TTF)$_2$KHg(SCN)$_4$}
\shorttitle{The angular dependent magnetoresistance\dots}
\author{Bal\'azs D\'ora\inst{1} \and Kazumi Maki\inst{2,3} \and Bojana
Korin-Hamzi\'c\inst{4} \and Mario Basletic\inst{5}  \and Attila
Virosztek\inst{1,6}\and Mark V. Kartsovnik\inst{7} \and Harald 
M\"uller\inst{8}}
\institute{
\inst{1} Department of Physics, Technical University of Budapest, H-1521 Budapest, Hungary \\
\inst{2} Max Planck Institute for the Physics of Complex Systems,
N\"othnitzer Str. 38, D-01187, Dresden, Germany\\
\inst{3} Department of Physics and Astronomy, University of Southern
California, Los Angeles
CA 90089-0484, USA \\
\inst{4} Institute of Physics, POB 304, HR-10001 Zagreb, Croatia\\
\inst{5} Department of Physics, Faculty of Science, POB 331, HR-10001
Zagreb, Croatia\\
\inst{6} Research Institute for Solid State Physics and Optics, P.O.Box
49,
H-1525 Budapest, Hungary\\
\inst{7} Walther-Meissner Institute, D-85748 Garching, Germany\\
\inst{8} European Synchrotron Radiation Facility, F-38043 Grenoble, 
France}
\shortauthor{Bal\'azs D\'ora \etal}


\date{}

\begin{document}

\maketitle

\begin{abstract}
In spite of extensive experimental studies of the angular dependent 
magnetoresistance (ADMR) of
the low temperature phase (LTP) of $\alpha$-(BEDT-TTF)$_2$KHg(SCN)$_4$ about a 
decade ago, the
nature of LTP remains elusive. Here we present a new study of ADMR of LTP 
in $\alpha-$(ET)$_2$
salts assuming that LTP is unconventional charge density wave (UCDW). In the
presence of magnetic field the quasiparticle spectrum in UCDW is quantized, 
which gives rise to
striking ADMR in UCDW. The present model appears to account for many existing 
ADMR data of
$\alpha$-(BEDT-TTF)$_2$KHg(SCN)$_4$ remarkably well.
 \end{abstract}

\section{Introduction}

The series of the quasi-two-dimensional organic conductors 
$\alpha$-(BEDT-TTF)$_2$MHg(SCN)$_4$ (where BEDT-TTF is 
bis(ethylenedithio)tetrathiafulvalene and M=K, NH$_4$, Rb and Tl) have 
attracted considerable attention in the last few years due to the richness 
of physical phenomena observed \cite{singl}. Whereas the M=NH$_4$ compound becomes 
superconducting below $1$ K, the other salts, at $T_c=8-10$ K, undergo a phase 
transition  into a specific low temperature phase (LTP), with associated 
numerous anomalies in magnetic field. LTP is thought to be caused by a 
density wave formation, but its nature appears still to be unsettled.
 We have proposed recently that unconventional charge density wave can
account for a number of features of LTP in
$\alpha$-(BEDT-TTF)$_2$KHg(SCN)$_4$ including the threshold
electric field\cite{kuszobter,rapid,tesla}. Recently,
 unconventional charge density wave (UCDW) and unconventional spin density
wave (USDW) have been proposed by several people as possible electronic
ground states in
quasi-one-dimensional and quasi-two-dimensional crystals\cite{Ners1,Ners2,benfatto,nagycikk,nayak}. Unlike conventional density wave\cite{gruner} the
order parameter in UCDW $\Delta(\bf k)$ depends on the quasiparticle wavevector $\bf k$.
In $\alpha$-(ET)$_2$ salts where the conducting plane lies in the a-c
plane and the quasi-one-dimensional Fermi surface is perpendicular to
the a-axis, we assume that $\Delta({\bf k})=\Delta\cos(ck_z)$ ( i.e. 
$\Delta(\bf k)$ depends on $\bf k$ perpendicular to the most conducting direction), 
where $c=9.778$ \AA $ $
is the lattice constant along the c axis\cite{endo}. It is known also that the 
thermodynamics of UCDW and USDW is identical to the one in d-wave superconductors\cite{nagycikk,d-wave}.
 Also, in spite of the clear thermodynamic signal, the first order term 
in $\Delta(\bf k)$ usually vanishes when averaged over the Fermi surface. This implies no clear X-ray or 
spin signal for UCDW and USDW. Due to this fact unconventional density wave ( i.e. UCDW and USDW) is sometimes
called the phase with hidden order parameter\cite{nayak}. In a magnetic field the quasiparticle spectrum is quantized
as first shown by Nersesyan et al.\cite{Ners1,Ners2}. This dramatic change in the quasiparticle spectrum is most readily seen
in ADMR, as demonstrated recently for SDW plus USDW in (TMTSF)$_2$PF$_6$ below $T=T^*$($\sim 4$K)\cite{romamaki,makitmtsf}.

About a decade ago ADMR in LTP in $\alpha$-(ET)$_2$ salts has been studied intensively. However the origin
of ADMR has been hotly debated\cite{fermi,caulfield1,caulfield2,hanasaki}. At that time the Fermi surface reconstruction due to nesting has been accepted as the most likely solution\cite{hibas,blundell}.

In the following we shall show that the quasiparticle spectrum in UCDW in $\alpha$-(ET)$_2$ salts is quantized
in the presence of magnetic field. The small energy gap which depends on both the direction and the strength 
of the magnetic field can be seen through ADMR. Indeed we can describe most aspects of ADMR seen in LTP of $\alpha$-(BEDT-TTF)$_2$KHg(SCN)$_4$ consistently. Therefore we may conclude that ADMR in $\alpha$-(BEDT-TTF)$_2$KHg(SCN)$_4$ 
provides conclusive evidence that the LTP is UCDW.

\section{Landau quantization}

\begin{figure}[h!]
\psfrag{B}[t][b][2][0]{$\bf B$}
\psfrag{c}[][][2][0]{$c$}
\psfrag{a}[][][2][0]{$a$}
\psfrag{b}[r][l][2][0]{$b$}
\psfrag{pp}[][][2][0]{$\phi$}
\psfrag{p}[][][2][0]{$\theta$}
\onefigure[width=7cm,height=7cm]{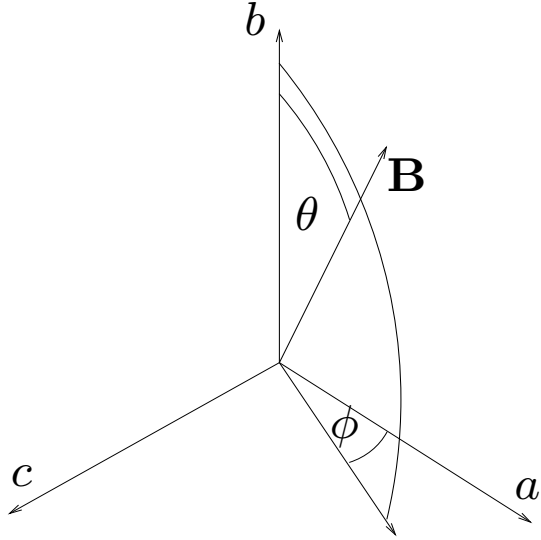}

\caption{The geometrical configuration of the magnetic field with respect to
 the conducting plane.}\label{fig:koord}
\end{figure}

We shall consider the configuration shown in Fig. \ref{fig:koord}, 
where a magnetic field 
$\bf B$ is applied within the a-b plane (i.e. $\phi=0$).
by angle $\theta$ tilted away from the b axis and in the 
plane with angle $\phi$ from the a axis. For simplicity we shall limit our 
analysis for $\phi=0$. Here the conducting plane is the a-c plane and the 
quasi-one-dimensional Fermi surfaces lie perpendicular to the a axis. In 
addition there is a quasi-two-dimensional Fermi surface with elliptical 
cross-section in the a-c plane.

Therefore we assume that there are two conducting channels in this system: 
the quasi-two-dimensional one stays
in the normal state while the quasi-one-dimensional one undergoes UCDW 
transition around $T=8$K. Then the 
quasiparticle spectrum in the quasi-1D channel is given by
\begin{equation}
E({\bf k})=\sqrt{\xi^2+\Delta({\bf k})^2}-\varepsilon_0\cos(2\bf b^\prime k),\label{elso}
\end{equation}
where $\xi\approx v_a k_a$, $\Delta({\bf k})=\Delta\cos(ck_c)$ and
$\varepsilon_0$ is the parameter describing
the imperfect nesting\cite{yamaji1,yamaji2,huang}. Finally $\bf b^\prime$ 
is the vector lying outside of the a-c plane. 
In order to fit the dip in ADMR at $\theta=\theta_0$ it is necessary to
tilt $\bf b^\prime$ from the b axis
by an angle $\theta_0$ \cite{hanasaki}. As is seen from Eq. (\ref{elso}),
the quasiparticle 
spectrum in the absence of magnetic field is gapless. When the magnetic field 
tilted by an angle $\theta$ from the b axis is applied, the lowest Landau level
above the Fermi surface is given by 
\begin{equation}
E(B,\theta)=\sqrt{2v_a\Delta c e
|B\cos\theta|}-\varepsilon_0\exp\left[-\frac{2\Delta c b^\prime}{v_a}e |B| 
\frac{\sin^2(\theta-\theta_0)}{|cos{\theta}|}\right].
\label{ketto}
\end{equation}
The first term in Eq. (\ref{ketto}) is obtained following Nersesyan et al. 
\cite{Ners1,Ners2}, while the second term in Eq. 
(\ref{ketto}) comes from the spatial average of the second term in Eq. 
(\ref{elso}) using the wavefunction of the Landau level at the Fermi
surface ($\sim exp(-v_a e |B\cos\theta|z^2/2c\Delta)$).

Noting the fact that the system has two conducting channels ( i.e.
quasi-1D Fermi surface and quasi-2D Fermi surface) and that only the
quasi-1D Fermi surface is affected by the formation of UCDW, the ADMR is
written as \begin{eqnarray}
R(B,\theta)=\dfrac{1}{\dfrac{4\sigma_1}{1+e^x}+\sigma_2},\label{harom}
\\
x=\beta E(B,\theta),
\end{eqnarray}
where only the thermal excitations to the lowest Landau level are taken into 
account explicitly, which is doubly degenerated.

Similarly we obtain
\begin{equation}
\frac{\Delta R}{R(0,0)}=\frac{2\sigma_1(e^x-1)}{[4\sigma_1+\sigma_2(1+e^x)]}.
\end{equation}

\section{Comparison with experiments}

\begin{figure}[h!]
\psfrag{xx}[t][b][1][0]{$T(K)$}
\psfrag{yy}[b][t][1][0]{$\Delta R/R(0,0)$} 
\psfrag{5T}[][][1][0]{$B=5$T}
\psfrag{x}[t][b][1][0]{$B(T)$}
\psfrag{y}[b][t][1][0]{$\Delta R/R(0,0)$} 
\psfrag{2.2K}[r][l][1][0]{$T=2.2$K}
\psfrag{4.2K}[l][r][1][0]{$T=4.2$K}
\twofigures[width=7cm,height=7cm]{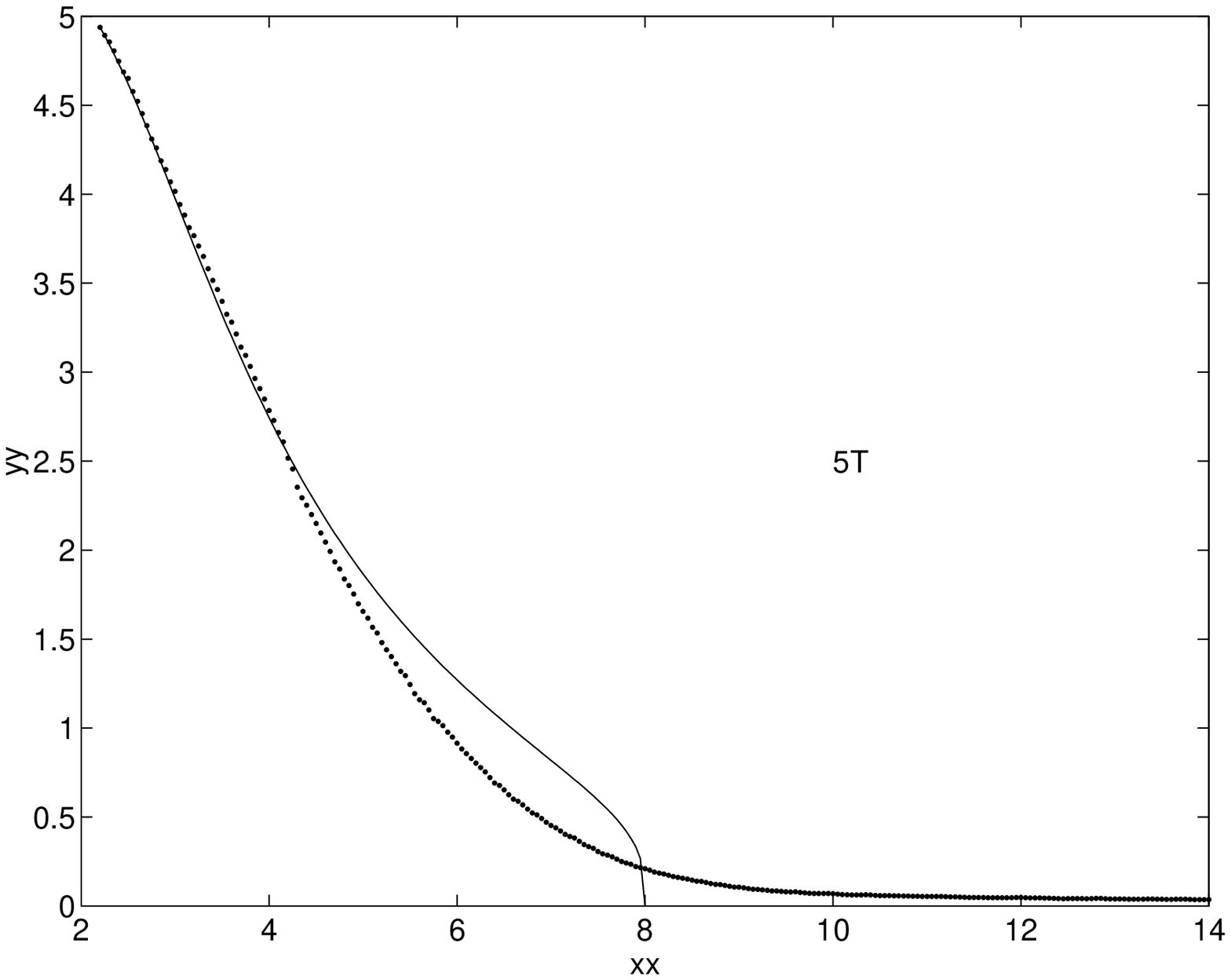}{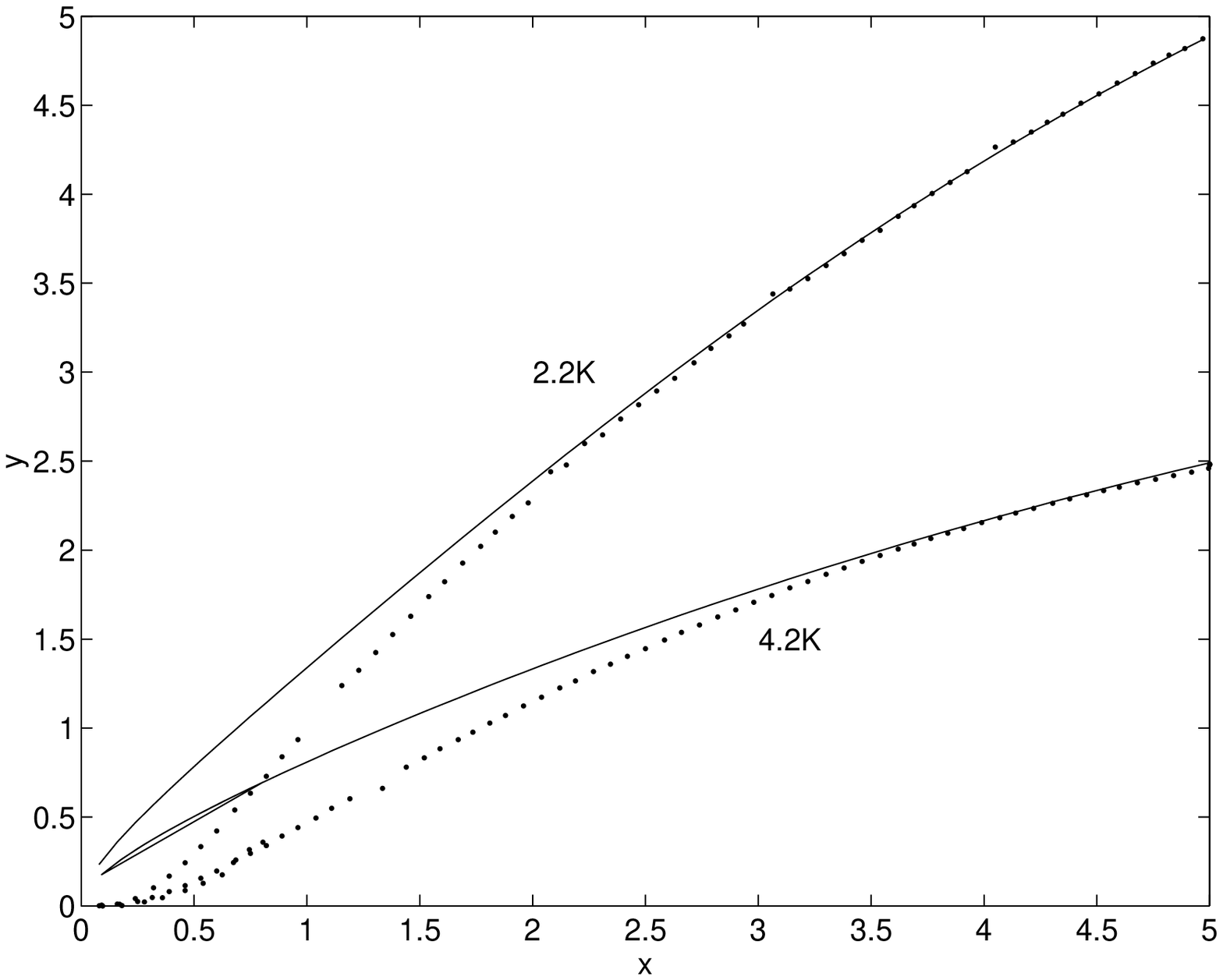}

\caption{The temperature dependence of the magnetoresistance is plotted at 
$B=5$T.}\label{fig:mrtemp}
\caption{The magnetic field dependence of the magnetoresistance is shown at 
$T=2.2$K and $4.2$K.}\label{fig:mrfield}
\end{figure}

First, we compare Eq. (\ref{harom}) with $R(B,\theta)$ versus $T$ and 
$R(B.\theta)$ versus $B$ in Figs. \ref{fig:mrtemp} and \ref{fig:mrfield}.
The temperature dependence of the magnetoresistance, for B=5T 
perpendicular to the a-c plane, is presented in Fig \ref{fig:mrtemp}. 
Solid line is the fit based on Eq. (\ref{harom}).
At low temperatures the consideration of the lowest 
Landau level provides convincing agreement. But the higher the temperature the
higher Landau levels should be taken into account, and close to $T_c$ the 
thermal
fluctuations play also an important role what we neglected here for simplicity. 
The strength of the two conducting channels was found to be 
$\sigma_2/\sigma_1=0.372$, 
and by assuming the weak coupling value of
$\Delta=17K$ and using $c=9.778$\AA \cite{endo}, 
the Fermi velocity is obtained as $v_a=7\times 10^6$cm/s.

The magnetic field dependence of the magnetoresistance at $T=2.2$ K and 
$T=4.2$ K for magnetic field perpendicular to the a-c plane is shown in 
Fig. \ref{fig:mrfield}. Solid line is the fit based on the Eq. (\ref{harom}).
At higher fields, where our simple
approximation is valid, the agreement looks perfect again. Here
$\sigma_2/\sigma_1=0.24$ and $0.48$ for $T=2.2$K and $4.2$K, respectively, and 
the extracted Fermi velocities (assuming again the weak coupling value of 
$\Delta$) are $v_a=3\times 10^6$ cm/s and 
$7\times 10^6$cm/s.

\begin{figure}[h!]
\psfrag{x}[t][b][1][0]{$\theta$}
\psfrag{y}[b][t][1][0]{$R(5T,\theta)$ $(\Omega)$} 
\psfrag{s}[l][r][1][0]{$T=4.2$K, $B=5$T}
\onefigure[width=7cm,height=7cm]{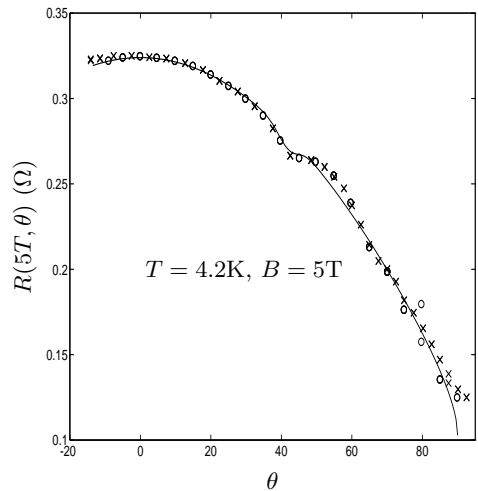}
\caption{The angle dependent magnetoresistance is shown at $T=4.2$K and $B=5$T.}\label{fig:mrtheta}
\end{figure}

Finally ADMR 
is shown in Fig. \ref{fig:mrtheta} as a function of angle $\theta$ at $T=4.2$K, 
$B=5$T. The solid line shows the fit to the theoretical model explained above.
As is seen from the figure 
the global $\theta$ dependence is given by $x\approx \beta
\sqrt{2v_a \Delta ce|B\cos\theta|}$, since the data is taken at $T=4.2$K and 
$B=5$T. 

The ratio of the conductivities in the two channels is $\sigma_2/\sigma_1=0.36$.
The different values of this ratio might arise from the fact that we considered
only the explicit $B$ and $T$ dependence of the main mechanism coming from 
the UCDW condensate, which according to us is responsible for the general
 behaviour of the measured
magnetoresistance, and we neglected the magnetic field 
and temperature dependence of the other possible conducting processes.
The Fermi velocity is obtained as $v_a=5\times 10^6$cm/s. From these
data, the Fermi velocity turned out to be of the order of $10^6$cm/s, and
its 
uncertainty should also be affected by the exclusion of the other conducting 
mechanisms.

The dip structure in ADMR is described fairly well by assuming 
$\varepsilon_0=0.132$K and $\theta_0=40^\circ$. This $\varepsilon_0$ 
value is an order of magnitude smaller than the one we needed to describe the 
threshold electric field observed in $\alpha$-(BEDT-TTF)$_2$KHg(SCN)$_4$
\cite{kuszobter,rapid,tesla,physicaB}. 
However, compared to other similar data (see \cite{hanasaki} ), it appears that there are 
considerable variability in the magnitude of $\varepsilon_0$ and the angle 
$\theta_0$ in different 
crystals ($\theta$ varies between $35^\circ$ and $50^\circ$). 
Therefore, we may conclude that UCDW in a magnetic field describes 
a variety of features in ADMR in $\alpha$-(BEDT-TTF)$_2$KHg(SCN)$_4$ 
satisfactorily.

\section{Concluding remarks}

From the analysis of the temperature dependence of the threshold electric field we have concluded earlier
that the LTP in $\alpha$-(ET)$_2$ salts is most likely
UCDW\cite{rapid}. The present analysis of ADMR appears to confirm this
identification. 
The quasiparticle spectrum in UCDW in magnetic field is quantized in
general\cite{Ners1,Ners2}. 
This effect should be most readily seen by the angular dependent
magnetoresistance as we have demonstrated in
(TMTSF)$_2$PF$_6$\cite{romamaki,makitmtsf}. 
We believe that ADMR will provide a powerful technique to explore other possible UCDW or USDW states in transition metal trichalogenate\cite{castroneto} and URu$_2$Si$_2$\cite{IO,roma} for example.

\section{Acknowledgements}

We thank Peter Thalmeier, Amir Hamzi\'c
and Silvia Tomi\'c for useful discussions on the related topics.
One of the authors (B. D.) gratefully
acknowledges the hospitality of the Max Planck Institute for the
Physics of Complex Systems, Dresden, where part of this work was done.
This work
was supported by the Hungarian National Research Fund under grant numbers
OTKA T032162 and T037451, and by the Ministry of Education under grant 
number FKFP 0029/1999.

\bibliographystyle{apsrev} 
\bibliography{mr}

\begin{thebibliography}{10}
\expandafter\ifx\csname bibnamefont\endcsname\relax
  \def\bibnamefont#1{#1}\fi
\expandafter\ifx\csname bibfnamefont\endcsname\relax
  \def\bibfnamefont#1{#1}\fi
\expandafter\ifx\csname url\endcsname\relax
  \def\url#1{\texttt{#1}}\fi
\expandafter\ifx\csname urlprefix\endcsname\relax\def\urlprefix{URL }\fi
\providecommand{\bibinfo}[2]{#2}
\providecommand{\eprint}[2][]{\url{#2}}

\bibitem{singl}
\bibinfo{author}{\bibnamefont{{s}ee for a~review J.~Singleton}},
  \bibinfo{journal}{Rep. Prog. Phys.} \textbf{\bibinfo{volume}{63}},
  \bibinfo{pages}{1161} (\bibinfo{year}{2000}).

\bibitem{kuszobter}
\bibinfo{author}{\bibfnamefont{M.}~\bibnamefont{Basleti{\'c}}},
  \bibinfo{author}{\bibfnamefont{B.}~\bibnamefont{Korin-Hamzi{\'c}}},
  \bibinfo{author}{\bibfnamefont{M.~V.} \bibnamefont{Kartsovnik}},
  \bibnamefont{and}
  \bibinfo{author}{\bibfnamefont{H.}~\bibnamefont{M{\"u}ller}},
  \bibinfo{journal}{Synth. Met.} \textbf{\bibinfo{volume}{120}},
  \bibinfo{pages}{1021} (\bibinfo{year}{2001}).

\bibitem{rapid}
\bibinfo{author}{\bibfnamefont{B.}~\bibnamefont{D\'ora}},
  \bibinfo{author}{\bibfnamefont{A.}~\bibnamefont{Virosztek}},
  \bibnamefont{and} \bibinfo{author}{\bibfnamefont{K.}~\bibnamefont{Maki}},
  \bibinfo{journal}{Phys. Rev. B} \textbf{\bibinfo{volume}{64}},
  \bibinfo{pages}{041101(R)} (\bibinfo{year}{2001}).

\bibitem{tesla}
\bibinfo{author}{\bibfnamefont{B.}~\bibnamefont{D\'ora}},
  \bibinfo{author}{\bibfnamefont{A.}~\bibnamefont{Virosztek}},
  \bibnamefont{and} \bibinfo{author}{\bibfnamefont{K.}~\bibnamefont{Maki}},
  \bibinfo{note}{{P}hys. Rev. B (in press)}.

\bibitem{Ners1}
\bibinfo{author}{\bibfnamefont{A.~A.} \bibnamefont{Nersesyan}}
  \bibnamefont{and} \bibinfo{author}{\bibfnamefont{G.~E.}
  \bibnamefont{Vachnadze}}, \bibinfo{journal}{J. Low Temp. Phys.}
  \textbf{\bibinfo{volume}{77}}, \bibinfo{pages}{293} (\bibinfo{year}{1989}).

\bibitem{Ners2}
\bibinfo{author}{\bibfnamefont{A.~A.} \bibnamefont{Nersesyan}},
  \bibinfo{author}{\bibfnamefont{G.~I.} \bibnamefont{Japaridze}},
  \bibnamefont{and} \bibinfo{author}{\bibfnamefont{I.~G.}
  \bibnamefont{Kimeridze}}, \bibinfo{journal}{J. Phys. Cond. Mat.}
  \textbf{\bibinfo{volume}{3}}, \bibinfo{pages}{3353} (\bibinfo{year}{1991}).

\bibitem{benfatto}
\bibinfo{author}{\bibfnamefont{L.}~\bibnamefont{Benfatto}},
  \bibinfo{author}{\bibfnamefont{S.}~\bibnamefont{Caprara}}, \bibnamefont{and}
  \bibinfo{author}{\bibfnamefont{C.}~\bibnamefont{{Di Castro}}},
  \bibinfo{journal}{Eur. Phys. J. B} \textbf{\bibinfo{volume}{17}},
  \bibinfo{pages}{95} (\bibinfo{year}{2000}).

\bibitem{nagycikk}
\bibinfo{author}{\bibfnamefont{B.}~\bibnamefont{D{\'o}ra}} \bibnamefont{and}
  \bibinfo{author}{\bibfnamefont{A.}~\bibnamefont{Virosztek}},
  \bibinfo{journal}{Eur. Phys. J. B} \textbf{\bibinfo{volume}{22}},
  \bibinfo{pages}{167} (\bibinfo{year}{2001}).

\bibitem{nayak}
\bibinfo{author}{\bibfnamefont{S.}~\bibnamefont{Chakravarty}},
  \bibinfo{author}{\bibfnamefont{R.~B.} \bibnamefont{Laughlin}},
  \bibinfo{author}{\bibfnamefont{D.~K.} \bibnamefont{Morr}}, \bibnamefont{and}
  \bibinfo{author}{\bibfnamefont{C.}~\bibnamefont{Nayak}},
  \bibinfo{journal}{Phys. Rev. B} \textbf{\bibinfo{volume}{63}},
  \bibinfo{pages}{094503} (\bibinfo{year}{2001}).

\bibitem{gruner}
\bibinfo{author}{\bibfnamefont{G.}~\bibnamefont{Gr\"uner}},
  \emph{\bibinfo{title}{Density waves in solids}}
  (\bibinfo{publisher}{Addison-Wesley}, \bibinfo{address}{Reading},
  \bibinfo{year}{1994}).

\bibitem{endo}
\bibinfo{author}{\bibfnamefont{S.}~\bibnamefont{Endo}},
  \bibinfo{author}{\bibfnamefont{Y.}~\bibnamefont{Watanabe}},
  \bibinfo{author}{\bibfnamefont{T.}~\bibnamefont{Sasaki}},
  \bibinfo{author}{\bibfnamefont{T.}~\bibnamefont{Fukase}}, \bibnamefont{and}
  \bibinfo{author}{\bibfnamefont{N.}~\bibnamefont{Toyota}},
  \bibinfo{journal}{Synth. Met.} \textbf{\bibinfo{volume}{86}},
  \bibinfo{pages}{2013} (\bibinfo{year}{1997}).

\bibitem{d-wave}
\bibinfo{author}{\bibfnamefont{H.}~\bibnamefont{Won}} \bibnamefont{and}
  \bibinfo{author}{\bibfnamefont{K.}~\bibnamefont{Maki}},
  \bibinfo{journal}{Phys. Rev. B} \textbf{\bibinfo{volume}{49}},
  \bibinfo{pages}{1397} (\bibinfo{year}{1994}).

\bibitem{romamaki}
\bibinfo{author}{\bibfnamefont{B.}~\bibnamefont{Korin-Hamzi\'c}},
  \bibinfo{author}{\bibfnamefont{M.}~\bibnamefont{Basletic}}, \bibnamefont{and}
  \bibinfo{author}{\bibfnamefont{K.}~\bibnamefont{Maki}},
  \bibinfo{note}{submitted to Int. J. Mod. Phys. B}.

\bibitem{makitmtsf}
\bibinfo{author}{\bibfnamefont{M.}~\bibnamefont{Basletic}},
  \bibinfo{author}{\bibfnamefont{B.}~\bibnamefont{Korin-Hamzi\'c}},
  \bibnamefont{and} \bibinfo{author}{\bibfnamefont{K.}~\bibnamefont{Maki}},
  \bibinfo{note}{submitted to Phys. Rev. B}.

\bibitem{fermi}
\bibinfo{author}{\bibfnamefont{T.}~\bibnamefont{Sasaki}} \bibnamefont{and}
  \bibinfo{author}{\bibfnamefont{N.}~\bibnamefont{Toyota}},
  \bibinfo{journal}{Phys. Rev. B} \textbf{\bibinfo{volume}{49}},
  \bibinfo{pages}{10120} (\bibinfo{year}{1994}).

\bibitem{caulfield1}
\bibinfo{author}{\bibfnamefont{J.}~\bibnamefont{Caulfield}},
  \bibinfo{author}{\bibfnamefont{S.~J.} \bibnamefont{Blundell}},
  \bibinfo{author}{\bibfnamefont{M.~S.~L.} \bibnamefont{du~Croo~de Jongh}},
  \bibinfo{author}{\bibfnamefont{P.~T.~J.} \bibnamefont{Hendriks}},
  \bibinfo{author}{\bibfnamefont{J.}~\bibnamefont{Singleton}},
  \bibinfo{author}{\bibfnamefont{M.}~\bibnamefont{Doporto}},
  \bibinfo{author}{\bibfnamefont{F.~L.} \bibnamefont{Pratt}},
  \bibinfo{author}{\bibfnamefont{A.}~\bibnamefont{House}},
  \bibinfo{author}{\bibfnamefont{J.~A. A.~J.} \bibnamefont{Perenboom}},
  \bibinfo{author}{\bibfnamefont{W.}~\bibnamefont{Hayes}},
  \bibinfo{author}{\bibfnamefont{M.}~\bibnamefont{Kurmoo}}, \bibnamefont{and}
  \bibinfo{author}{\bibfnamefont{P.}~\bibnamefont{Day}},
  \bibinfo{journal}{Phys. Rev. B} \textbf{\bibinfo{volume}{51}},
  \bibinfo{pages}{8325} (\bibinfo{year}{1995}).

\bibitem{caulfield2}
\bibinfo{author}{\bibfnamefont{J.}~\bibnamefont{Caulfield}},
  \bibinfo{author}{\bibfnamefont{J.}~\bibnamefont{Singleton}},
  \bibinfo{author}{\bibfnamefont{P.~T.~J.} \bibnamefont{Hendriks}},
  \bibinfo{author}{\bibfnamefont{J.~A. A.~J.} \bibnamefont{Perenboom}},
  \bibinfo{author}{\bibfnamefont{F.~L.} \bibnamefont{Pratt}},
  \bibinfo{author}{\bibfnamefont{M.}~\bibnamefont{Doporto}},
  \bibinfo{author}{\bibfnamefont{W.}~\bibnamefont{Hayes}},
  \bibinfo{author}{\bibfnamefont{M.}~\bibnamefont{Kurmoo}}, \bibnamefont{and}
  \bibinfo{author}{\bibfnamefont{P.}~\bibnamefont{Day}}, \bibinfo{journal}{J.
  Phys. Cond. Mat.} \textbf{\bibinfo{volume}{6}}, \bibinfo{pages}{L155}
  (\bibinfo{year}{1994}).

\bibitem{hanasaki}
\bibinfo{author}{\bibfnamefont{N.}~\bibnamefont{Hanasaki}},
  \bibinfo{author}{\bibfnamefont{S.}~\bibnamefont{Kagoshima}},
  \bibinfo{author}{\bibfnamefont{N.}~\bibnamefont{Miura}}, \bibnamefont{and}
  \bibinfo{author}{\bibfnamefont{G.}~\bibnamefont{Saito}}, \bibinfo{journal}{J.
  Phys. Soc. Japan} \textbf{\bibinfo{volume}{65}}, \bibinfo{pages}{1010}
  (\bibinfo{year}{1996}).

\bibitem{hibas}
\bibinfo{author}{\bibfnamefont{M.~V.} \bibnamefont{Kartsovnik}},
  \bibinfo{author}{\bibfnamefont{A.~E.} \bibnamefont{Kovalev}},
  \bibnamefont{and} \bibinfo{author}{\bibfnamefont{N.~D.}
  \bibnamefont{Kushch}}, \bibinfo{journal}{J. Phys. I France}
  \textbf{\bibinfo{volume}{3}}, \bibinfo{pages}{1187} (\bibinfo{year}{1993}).

\bibitem{blundell}
\bibinfo{author}{\bibfnamefont{S.~J.} \bibnamefont{Blundell}} \bibnamefont{and}
  \bibinfo{author}{\bibfnamefont{J.}~\bibnamefont{Singleton}},
  \bibinfo{journal}{Phys. Rev. B} \textbf{\bibinfo{volume}{53}},
  \bibinfo{pages}{5609} (\bibinfo{year}{1996}).

\bibitem{yamaji1}
\bibinfo{author}{\bibfnamefont{K.}~\bibnamefont{Yamaji}}, \bibinfo{journal}{J.
  Phys. Soc. Japan} \textbf{\bibinfo{volume}{51}}, \bibinfo{pages}{2787}
  (\bibinfo{year}{1982}).

\bibitem{yamaji2}
\bibinfo{author}{\bibfnamefont{K.}~\bibnamefont{Yamaji}}, \bibinfo{journal}{J.
  Phys. Soc. Japan} \textbf{\bibinfo{volume}{52}}, \bibinfo{pages}{1361}
  (\bibinfo{year}{1983}).

\bibitem{huang}
\bibinfo{author}{\bibfnamefont{X.~Z.} \bibnamefont{Huang}} \bibnamefont{and}
  \bibinfo{author}{\bibfnamefont{K.}~\bibnamefont{Maki}},
  \bibinfo{journal}{Phys. Rev. B} \textbf{\bibinfo{volume}{40}},
  \bibinfo{pages}{2575} (\bibinfo{year}{1989}).

\bibitem{physicaB}
\bibinfo{author}{\bibfnamefont{B.}~\bibnamefont{D{\'o}ra}},
  \bibinfo{author}{\bibfnamefont{A.}~\bibnamefont{Virosztek}},
  \bibnamefont{and} \bibinfo{author}{\bibfnamefont{K.}~\bibnamefont{Maki}},
  \bibinfo{note}{{P}hysica B (in press)}.

\bibitem{castroneto}
\bibinfo{author}{\bibfnamefont{A.~H.} \bibnamefont{Castro-Neto}},
  \bibinfo{journal}{Phys. Rev. Lett.} \textbf{\bibinfo{volume}{86}},
  \bibinfo{pages}{4382} (\bibinfo{year}{2001}).

\bibitem{IO}
\bibinfo{author}{\bibfnamefont{H.}~\bibnamefont{Ikeda}} \bibnamefont{and}
  \bibinfo{author}{\bibfnamefont{Y.}~\bibnamefont{Ohasi}},
  \bibinfo{journal}{Phys. Rev. Lett.} \textbf{\bibinfo{volume}{81}},
  \bibinfo{pages}{3723} (\bibinfo{year}{1998}).

\bibitem{roma}
\bibinfo{author}{\bibfnamefont{A.}~\bibnamefont{Virosztek}},
  \bibinfo{author}{\bibfnamefont{K.}~\bibnamefont{Maki}}, \bibnamefont{and}
  \bibinfo{author}{\bibfnamefont{B.}~\bibnamefont{D{\'o}ra}},
  \bibinfo{note}{cond-mat/0112387, to be published in Int. J. Mod. Phys. B}.

\end{thebibliography}
\end{document}